\documentclass[conference]{IEEEtran}
\usepackage{amsmath, amssymb, bm, cite, epsfig, psfrag, mathtools}
\usepackage{epstopdf}
\usepackage{graphicx}
\usepackage{algorithm,algpseudocode}
\usepackage{array}
\usepackage[margin=10pt,font=small]{caption}
\usepackage{bbm}
\usepackage{multirow}
\usepackage{fancybox}
\usepackage[usenames,dvipsnames]{xcolor}

\usepackage{subfigure}
\usepackage{etoolbox}
\usepackage{pbox}
\usepackage{tikz}
\usetikzlibrary{calc}

\newtoggle{conference}
\graphicspath{{figures/}}

\def\beq{\begin{equation}}
\def\eeq{\end{equation}}
\def\beqa{\begin{eqnarray}}
\def\eeqa{\end{eqnarray}}
\def\beqan{\begin{eqnarray*}}
\def\eeqan{\end{eqnarray*}}

\def\PL{\mathrm{PL}}
\def\dB{\mathrm{dB}}

\def\tm1{t\! - \! 1}
\def\tp1{t\! + \! 1}

\begin{document}
\thispagestyle{empty}
\title{28 GHz and 73 GHz Millimeter-Wave Indoor Propagation Measurements and Path Loss Models }

\author{
	Sijia Deng,~\IEEEmembership{Student Member,~IEEE,}
    Mathew K. Samimi,~\IEEEmembership{Student Member,~IEEE,}
   	Theodore S. Rappaport,~\IEEEmembership{Fellow,~IEEE}

\IEEEauthorblockA
{\\NYU WIRELESS\\NYU Polytechnic School of Engineering, Brooklyn, NY 11201\\ sijia@nyu.edu, mks@nyu.edu, tsr@nyu.edu}
}

\maketitle

\begin{tikzpicture} [remember picture, overlay]
\node at ($(current page.north) + (0,-0.25in)$) {S. Deng, M. K. Samimi, T. S. Rappaport, ''28 GHz and 73 GHz Millimeter-Wave Indoor Propagation Measurements and Path Loss Models,''};
\node at ($(current page.north) + (0,-0.4in)$) {\textit{accepted at the 2015 IEEE International Conference on Communications Workshop (ICC Workshop)}, 8-12 June, 2015.};
\end{tikzpicture}

\thispagestyle{empty}

\begin{abstract}
This paper presents 28 GHz and 73 GHz millimeter-wave propagation measurements performed in a typical office environment using a 400 Megachip-per-second broadband sliding correlator channel sounder and highly directional steerable 15 dBi (30$^{\circ}$ beamwidth) and 20 dBi (15$^{\circ}$ beamwidth) horn antennas. Power delay profiles were acquired for 48 transmitter-receiver location combinations over distances ranging from 3.9 m to 45.9 m with maximum transmit powers of 24 dBm and 12.3 dBm at 28 GHz and 73 GHz, respectively. Directional and omnidirectional path loss models and RMS delay spread statistics are presented for line-of-sight and non-line-of-sight environments for both co- and cross-polarized antenna configurations. The LOS omnidirectional path loss exponents were 1.1 and 1.3 at 28 GHz and 73 GHz, and 2.7 and 3.2 in NLOS at 28 GHz and 73 GHz, respectively, for vertically-polarized antennas. The mean directional RMS delay spreads were 18.4 ns and 13.3 ns,  with maximum values of 193 ns and 288 ns at 28 GHz and 73 GHz, respectively. 

\end{abstract}


\section{Introduction}\label{sec:intro}

The overwhelming demand for broadband wireless communications is expected to increase by a factor of 10,000 over the next 10 years, which is motivating the use of the millimeter-wave (mmWave) spectrum where a vast amount of available raw bandwidth will provide multi-gigabit-per-second transmission throughputs to mobile devices for next generation 5G wireless system~\cite{TedNewBook,Nokia2014JSAC,B4GNSN}. The 28 and 38 GHz Local to Multipoint Distribution Service (LMDS) frequency bands and the E-band are serious candidates for mmWave communications, with more than 3 GHz and 10 GHz of available bandwidth, respectively \cite{ZPiIntro2011,Nokia2014JSAC}. The E-band (71-76 GHz, 81-86 GHz, and 92-95 GHz frequency bands) has recently become available to provide ultra-high-speed data communications in point-to-point wireless local area networks (WLANs), mobile backhaul, and broadband Internet access \cite{ZPiIntro2011}. While the E-band contains a massive amount of raw spectrum, it has so far received little research attention, thereby driving the community to study its propagation characteristics to extract statistical channel models~\cite{MicrowaveJ:DSM14}. Indoor channel measurements are vital to understand path loss as a function of distance, and temporal and spatial characteristics, which are crucial in performing system-wide simulations to estimate network system capacities and overall data throughputs.

Indoor wireless channels are currently served over 2.4 GHz, 5 GHz WiFi, and 60 GHz WiGig frequency bands, commonly used for short-range indoor communications. The vast available bandwidth (6 GHz) in the 60 GHz mmWave band has motivated extensive 60 GHz indoor propagation measurements to understand channel characteristics for designing WLAN systems, capable of achieving multi-gigabits-per-second throughputs~\cite{60GWLAN,HX60G,Droste60G}. Highly directional horn antennas have also been placed at the TX to overcome the additional 15 dB/km of atmospheric attenuation, while reducing inter-cell interference~\cite{60GWLAN}. Typical measured path loss exponents (PLEs) in indoor line-of-sight (LOS) environments were measured to be 1.3 in corridors, 1.7 in a laboratory~\cite{Tharek88}, and 2.2 in an office area~\cite{Myoung60G}. Larger PLEs in non-line-of-sight (NLOS) environments were reported, ranging from 3.0 to 3.8 in typical office environments~\cite{Myoung60G,Geng05}. Average  RMS delay spreads were 12.3 ns and 14.6 ns in LOS and NLOS environments at 60 GHz, respectively~\cite{Myoung60G}. 

 This paper presents extensive 28 GHz and 73 GHz mmWave indoor propagation measurements that can be used to extract omnidirectional and directional path loss channel models, and time dispersion characteristics to gain insight into the design of mmWave communication system in indoor environment.

\section{Millimeter-Wave Indoor Propagation Measurements}     

\subsection{Propagation Measurements and Environment Description}

\begin{figure*}
\centering
\includegraphics[width=\textwidth]{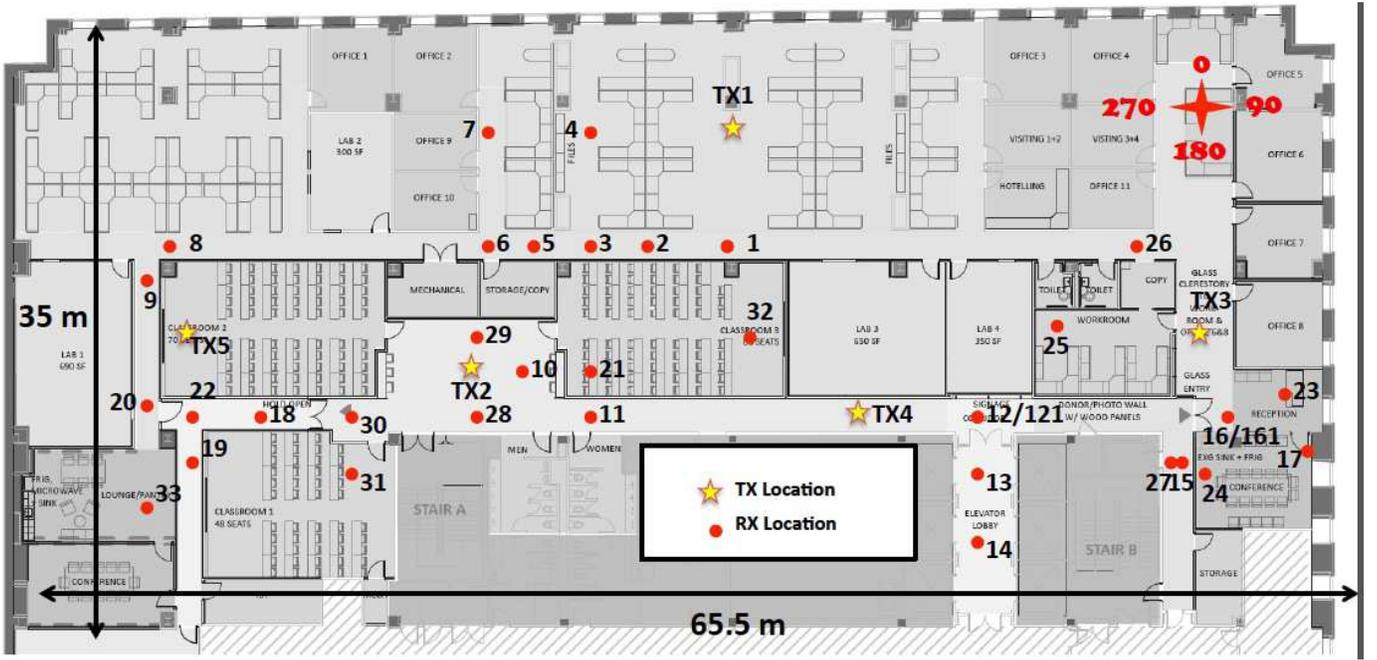}
\caption{Indoor map where 28 GHz and 73 GHz propagation measurements were conducted within a modern office building (65.5 m $\times$ 35 m $\times$ 2.7 m) at five TX locations and 33 RX locations, for a total of 48 TX-RX location combinations. The interior of the building included materials such as drywalls, glass doors, metal doors, elevator doors and soft office partitions. These extensive measurements can be used to extract accurate statistical channel models for mmWave indoor system design.}
\label{fig:Map}
\end{figure*}

Two indoor propagation measurement campaigns were conducted in a typical office environment at 28 GHz and 73 GHz using a 400 Megachip-per-second (Mcps) spread spectrum broadband sliding correlator channel sounder. Two pairs of 15 dBi (30$^{\circ}$ half-power beamwidth (HPBW)) and 20 dBi (15$^{\circ}$ HPBW) high gain directional antennas were employed at the TX and RX, and rotated exhaustively in the azimuth and elevation dimensions to recover AOD and AOA spatial statistics. The TX and RX were placed 2.5 m and 1.5 m above ground level, respectively, so as to emulate a typical WLAN network environment. Five TX locations and 33 RX locations were tested with transmitter-receiver (T-R) separation distances ranging from 3.9 m to 45.9 m in a typical office environment as shown in Fig.~\ref{fig:Map}. The measurements were conducted within a modern office building (65.5 m $\times$ 35 m $\times$ 2.7 m) with common office partitions (such as cubicles, desks, chairs, metal shelves, wood closets), concrete walls, glass doors and elevator doors. For each measured TX-RX location combination, eight different unique pointing angle measurement sweeps were performed at both the TX and RX to investigate angles of departure (AODs) and angles of arrival (AOAs) statistics, and a power delay profile (PDP) was acquired at each azimuth and elevation unique pointing angle in step increments of 15$^{\circ}$ or 30$^{\circ}$ depending on the carrier frequency. All azimuth sweeps were performed in both vertical-to-vertical (V-V) and vertical-to-horizontal (V-H) antenna polarization scenarios to study de-polarization effects.

Indoor channel propagation environments for each TX-RX location combination are categorized into LOS and NLOS, depending on whether there was an unobstructed path between the TX and RX antennas. When using omnidirectional antennas or omnidirectional models, LOS refers to a scenario where there was an unobstructed path between the TX and RX antenas, whereas NLOS refers to environment where there were obstructions between TX and RX. For directional antennas and directional models, LOS refers to a scenario where the TX and RX antennas were aligned on boresight with no obstructions between them (LOS boresight), while NLOS refers to a scenario where the TX and RX antennas were not aligned on boresight, regardless of whether the LOS path was obstructed or not (including LOS nonboresight and NLOS).

\subsection{Measurement Equipment}

Two 400 Mcps broadband sliding correlator channel sounders were employed at 28 GHz and 73 GHz that used similar architectures with varying Intermediate Frequency (IF) and Local Oscillator (LO) frequencies, as well as inter-changeable RF up- and down-converter front-ends. At the TX, a maximal length 2047 pseudorandom noise (PN) sequence clocked at 400 MHz, was upconverted to 5.4 GHz and 5.625 GHz at 28 GHz and 73 GHz respectively, and subsequently mixed with a 22.6 GHz and 67.875 GHz LO, allowing us to reach 28 GHz and 73.5 GHz carrier frequencies \cite{ShuaiPIMRC}, and finally fed through a high gain directional horn antenna. The received in-phase ($\textit{I}$) and quadrature ($\textit{I}$) signal components were obtained after downconversion from RF, and cross-correlated with an identical PN sequence clocked at a slightly lower rate of 399.95 MHz, providing a slide factor of 8,000. The final recorded PDPs were obtained by squaring and summing the $I$ and $Q$ channels, and time-averaging 20 instantaneous PDP measurements using real-time trigger alignment to the strongest measured multipath component. Table~\ref{tbl:Specs} shows the 28 GHz and 73 GHz system specifications, and more information can be found in \cite{5GItwillwork,GeorgeICC2014}.

\begin{table}
\centering
\caption{Broadband sliding correlator channel sounding system specifications used at 28 GHz and 73 GHz.}

\begin{tabular}{|c|p{1.5cm}|p{1.5cm}|}

\hline
\textbf{Carrier Frequency} &\centering \textbf{28 GHz} &\centering \textbf{73.5 GHz} \tabularnewline \hline

\textbf{RF Bandwidth (first null)} & \multicolumn{2}{c|}{800 MHz} \tabularnewline \hline

\textbf{Chip Sequence Length} & \multicolumn{2}{c|}{2047} \tabularnewline \hline

\textbf{TX Chip Rate} & \multicolumn{2}{c|}{400 Mcps} \tabularnewline \hline

\textbf{Slide Factor} & \multicolumn{2}{c|}{8000} \tabularnewline \hline

\textbf{TX/RX IF Frequency} & \centering 5.4 GHz & \centering 5.625 GHz \tabularnewline \hline

\textbf{TX/RX LO Frequency} & \centering 22.6 GHz & \centering 22.625 GHz \tabularnewline \hline


\textbf{Maximum TX Output Power} & \centering 24 dBm & \centering 14.6 dBm \tabularnewline \hline

\textbf{TX/RX Antenna Gains} & \centering 15 dBi & \centering 20 dBi \tabularnewline \hline

\textbf{TX/RX Azi. HPBW} & \centering 30$^{\circ}$ & \centering 15$^{\circ}$ \tabularnewline \hline

\textbf{TX/RX Elv. HPBW} & \centering 28.8$^{\circ}$ & \centering 15$^{\circ}$ \tabularnewline \hline

\textbf{Maximum Measurable Path Loss} & \centering 162 dB & \centering 163 dB \tabularnewline \hline

\textbf{Multipath Time Resolution} & \multicolumn{2}{c|}{2.5 ns} \tabularnewline \hline 

\textbf{TX-RX Synchronization} & \multicolumn{2}{c|}{{Unsupported}} \tabularnewline \hline

\end{tabular}
\label{tbl:Specs}
\end{table}

\subsection{Measurement Procedure}
  
At each TX-RX location, the TX and RX antennas were rotated exhaustively in azimuth and elevation to collect AOA and AOD statistics of the indoor wireless channel. We performed eight individual unique pointing azimuth sweeps at various elevation planes. For each azimuth sweep, we stepped the TX or RX antenna in 15$^{\circ}$ or 30$^{\circ}$ increments (depending on the carrier frequency) and acquired a PDP measurement for fixed TX and RX antenna positions.  The eight azimuth sweeps included one AOA sweep (Measurement1, M1) and one AOD sweep (M2) where the TX and RX antennas were perfectly aligned on boresight in the elevation planes. 

Additionally, two AOA azimuth sweeps with the RX antenna uptilted (M3) and downtilted (M4) by one antenna HPBW with respect to the boresight elevation angle with the TX antenna fixed at the boresight elevation angle, two AOA sweeps with the TX antenna uptilted (M5) and downtilted (M6) by one antenna HPBW with respect to the boresight elevation angle while the RX antenna remained fixed at the boresight elevation angle, and one AOA sweep (M7) with the TX antenna set to the second strongest AOD (obtained from M2) were conducted. Finally, a second AOD sweep (M8) with the TX antenna either uptilted or downtilted by one antenna HPBW was performed after determining the elevation plane with the strongest received power resulting from M5 and M6. These measurement sweeps were performed at each TX-RX location combination in both V-V and V-H antenna polarization configurations.

\section{Measurement Results and Analysis}

\subsection{Directional 28 GHz and 73 GHz Path Loss Models}

\begin{figure}
\centering
\includegraphics[width=3.5in]{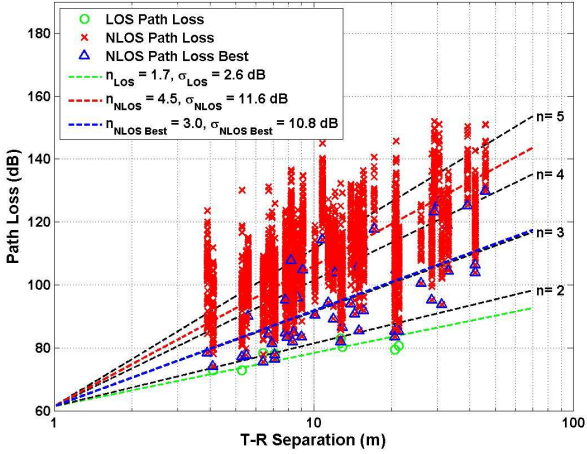}
\caption{28 GHz directional path loss models for vertically co-polarized TX and RX antennas using the 1 m close-in free space reference distance path loss model. Squares and circles represent LOS path loss values (meaning LOS boresight), crosses represent NLOS path loss values (including LOS non-boresight and NLOS), and triangles represent the smallest path loss values measured for a specific TX-RX location combination.}
\label{fig:Scatter_28_VV}
\end{figure}

\begin{figure}
\centering
\includegraphics[width=3.5in]{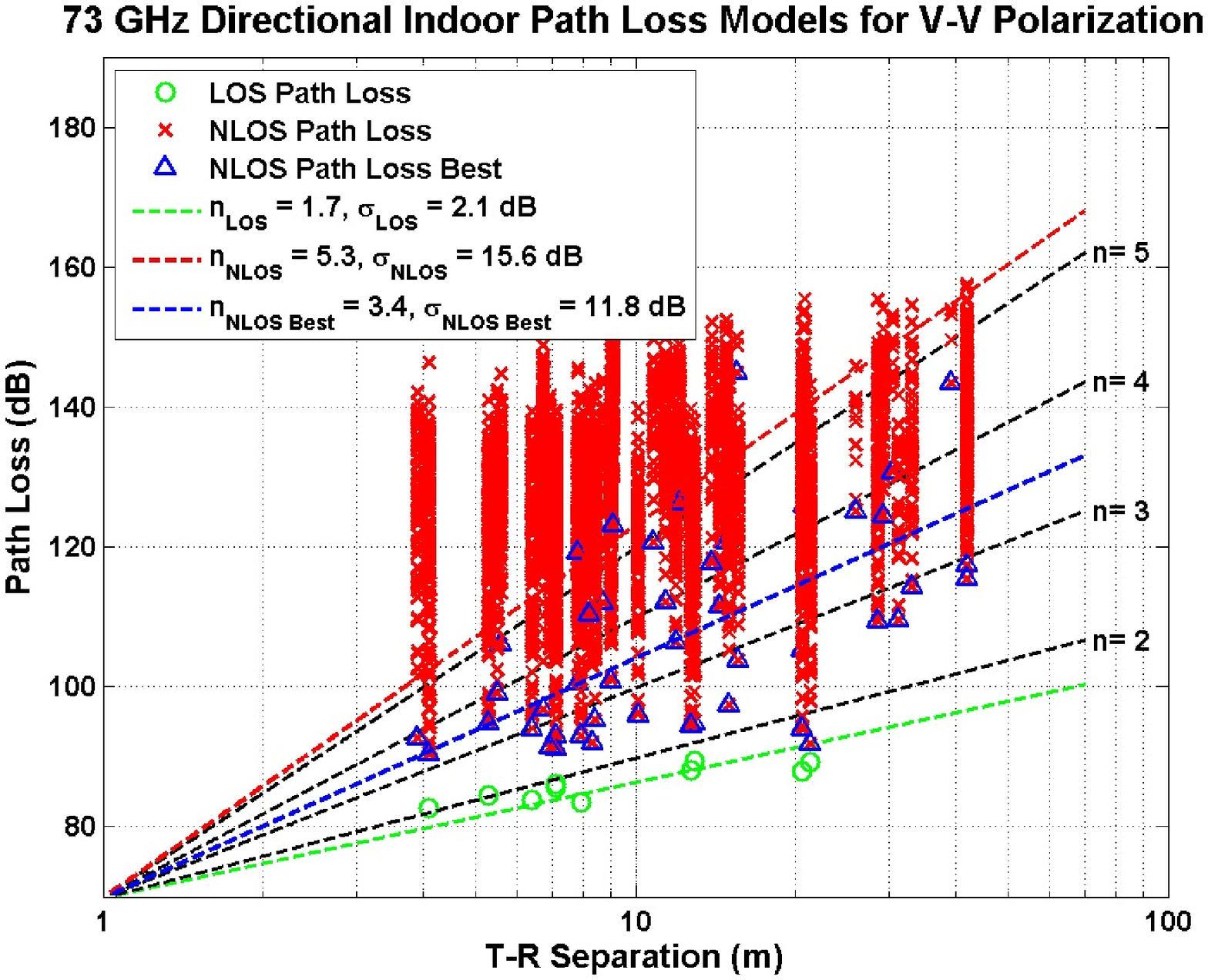}
\caption{73 GHz directional path loss models for vertically polarized TX and RX antennas using the 1 m close-in free space reference distance path loss model. Squares and circles represent LOS path loss values (meaning LOS boresight), crosses represent NLOS path loss values (including LOS non-boresight and NLOS), and triangles represent the smallest path loss values measured for a specific TX-RX location combination.}
\label{fig:Scatter_73_VV}
\end{figure}

The directional path loss models are useful in estimating path loss at arbitrary unique pointing angles in a mmWave communication channel, where measured signal power levels are very sensitive to TX and RX antenna pointing directions. The received power obtained at a TX-RX unique pointing angle combination was obtained by summing the power of each individual multipath component in time. The corresponding path loss was recovered by subtracting the TX power (in dBm) and removing TX and RX antenna gains. Fig.~\ref{fig:Scatter_28_VV} and Fig.~\ref{fig:Scatter_73_VV} show the LOS and NLOS directional path loss data, and corresponding 1 m close-in free space reference path loss equation lines obtained using the minimum mean square error (MMSE) fit, at both 28 GHz and 73 GHz, for the V-V polarization scenario. The measured LOS path loss exponent (PLE) for arbitrary unique pointing angles was 1.7, with shadowing factors of 2.6 dB and 2.1 dB, at both 28 GHz and 73 GHz, which are slightly less than the theoretical free space propagation exponent (n = 2), and most likely a result of the waveguide effect occuring from indoor hallways and partitions. In NLOS, the measured PLEs and shadowing factors were 4.5 and 11.6 dB, and 5.3 and 15.6 at 28 GHz and 73 GHz, respectively, indicating faster signal level degradation over distance. When considering the strongest TX-RX antenna pointing angle link at each TX-RX location combination, the PLEs drop to 3.0 and 3.4 at 28 GHz and 73 GHz, respectively, indicating the benefit of implementing beamforming at the mobile handset. Work in~\cite{ShuICC14} has also demonstrated significant reductions in path loss when performing both beamforming and beamcombining of multiple beams at the mobile handset. In cross-polarized V-H scenario, the measured PLEs and shadowing factors in NLOS were 5.1 and 10.9 dB, and 6.4 and 15.8 dB at 28 GHz and 73 GHz, respectively. Table~\ref{tbl:PLEVV} and Table~\ref{tbl:PLEVH} summarize the 28 GHz and 73 GHz measured PLEs and corresponding shadow factors ($\sigma$) in LOS and NLOS, and for V-V and V-H polarization scenarios.

\begin{table}
\centering
\caption{Summary of measured 28 GHz and 73 GHz PLEs and standard deviations for both directional and omnidirectional path loss models and for co-polarization V-V scenario.}
\label{tbl:PLEVV}
\begin{center}
\begin{tabular}{|c|c|c|c|c|c|}

\hline
\multirow{2}{*}{\textbf{Path Loss Model}}	& \multirow{2}{*}{\textbf{Scenario}} & \multicolumn{2}{c|}{\textbf{28 GHz}}  & \multicolumn{2}{c|}{\textbf{73 GHz}}   \\ \cline{3-6} 

						  	&						& \textbf{$\bar n$} & \textbf{$\sigma$} [dB] & \textbf{$\bar n$} & \textbf{$\sigma$ }[dB] \\   \hline

\multirow{2}{*}{\textbf{Directional}}			& \textbf{LOS} 	& 1.7 	& 2.6   & 1.7 	& 2.1    \\    \cline{2-6}
\multirow{2}{*}{\textbf{(V-V)}}				& \textbf{NLOS}       & 4.5 	& 11.6  & 5.3  	& 15.6   \\    \cline{2-6}
							 	& \textbf{NLOS-Best}  & 3.0  	& 10.8 	& 3.4   & 11.8   \\    \hline

\textbf{Omni.}						& \textbf{LOS}  	   & 1.1    & 1.7   & 1.3   & 1.9    \\    \cline{2-6}
\textbf{(V-V)}						& \textbf{NLOS} 	   & 2.7    & 9.6   & 3.2   & 11.3   \\    \hline

 \end{tabular}
\end{center}
\end{table}
 
\begin{table}
\centering
\caption{Summary of measured 28 GHz and 73 GHz PLEs and standard deviations for both directional and omnidirectional path loss models and for cross-polarization V-H scenario.}
 \label{tbl:PLEVH}
 \centering
 \begin{center}
  \begin{tabular}{|c|c|c|c|c|c|}
	
	 \hline
	 \multirow{2}{*}{ \textbf{Path Loss Model} }& \multirow{2}{*}{\textbf{Scenario}} & \multicolumn{2}{c|}{\textbf{28 GHz}}  & \multicolumn{2}{c|}{\textbf{73 GHz}}   \\ \cline{3-6} 
	  && \textbf{$\bar n$} & \textbf{$\sigma$} [dB] & \textbf{$\bar n$} & \textbf{$\sigma$} [dB] \\   \hline

	   \multirow{2}{*}{\textbf{Directional}} & \textbf{LOS}        	& 4.1 	& 8.0 	& 4.7 	& 9.0   \\   \cline{2-6}

	\multirow{2}{*}{\textbf{(V-H)}}                                         & \textbf{NLOS}        & 5.1   & 10.9  & 6.4   & 15.8  \\   \cline{2-6}
	                                         & \textbf{NLOS-Best}   & 4.3   & 9.1   & 5.0   & 10.9  \\   \hline

	    \textbf{Omni.} 		 & \textbf{LOS}   		& 2.5   & 3.0   & 3.5   & 6.3   \\   \cline{2-6}
	      \textbf{(V-H)}                                   & \textbf{NLOS} 	    & 3.6   & 9.4   & 4.6   & 9.7   \\   \hline

 \end{tabular}
\end{center}
\end{table}

\subsection{Omnidirectional Close-in Reference Path Loss Models}

Omnidirectional path loss models are required to estimate the total received power at a given T-R separation distance when performing system-wide simulations, using arbitrary antenna patterns. The omnidirectional received powers were synthesized by summing the directional received powers at each and every unique TX-RX azimuth and elevation pointing angle combination, which is a valid procedure because measured signal at each azimuth and elevation angle suffered little interference from adjacent bins, as a result of the one HPBW spacing between measured angles, offering a near orthogonal antenna pattern for each bin. Arrivng signals from adjacent bins travelled different propagation distances, so that the phase of individual multipath components can assumed to be uncorrelated, thus allowing powers of each resolvable multipath component to be summed over the omnidirectional spatial manifold ~\cite{TedBook,MKR:PIMRC14}. The corresponding omnidirectional path losses were recovered by subtracting the transmit power (in dBm) and removing antenna gains. The omnidirectional received power and path loss were obtained as~\cite{MKR:PIMRC14}:

\begin{equation}
\mathrm {Pr_{\mathrm{omni}}(d)} =  \sum_{i,j} \sum_{k,m} \mathrm{Pr}(\theta_{TX,i},\phi_{TX,j},\theta_{RX,k},\phi_{RX,m})
\end{equation}
\begin{equation}\label{E2}
\PL[\dB](d) = \mathrm{P_{TX}} + \mathrm{G_t} + \mathrm{G_r} - 10 \times \log_{10}(\mathrm {Pr_{\mathrm{omni}}})
\end{equation}

\noindent where $\theta_{TX}, \phi_{TX}, \theta_{RX}, \phi_{RX}$ denote the TX azimuth and elevation angles, and the RX azimuth and elevation angles, respectively, $i, j, k, m$ correspond to indices for each TX azimuth and elevation angles, and RX azimuth and elevation angles, $\mathrm {P_{TX}}$ is the transmit power in dBm, and $\mathrm{G_t}$ and $\mathrm{G_r}$ are the TX and RX antenna gains in dBi, respectively. All path losses were recovered and the corresponding close-in free space reference path loss model with respect to a 1 m free space reference distance was extracted by recovering the omnidirectional path loss exponents and shadow factors for different polarization scenarios in LOS and NLOS environments, at both 28 GHz and 73 GHz. 

The $d_0=1$ m close-in free space reference path loss model has the following form~\cite{TedBook}: 
\begin{equation}
\PL (d) [\dB]= \PL_{FS} (d_{0}) [\dB]+10 \cdot \bar n \cdot \log_{10}    \left(\frac{ d}{d_{0}}\right) + \chi_{\sigma}
\end{equation}
\begin{equation}
\PL_{FS} (d_{0}) [\dB]= 20  \cdot \log_{10} \left(  \frac{4 \pi d_0}{\lambda}   \right)
\end{equation}

\noindent where $\PL_{\mathrm{FS}} (d_{0})$ is the free space path loss at distance $d_0$, $\overline{n}$ is the omnidirectional path loss exponent obtained using the minimum mean square error method (MMSE), $\lambda$ is the carrier wavelength, and $\chi_{\sigma}$ is a 0 dB mean lognormal random variable with standard deviation $\sigma$ (also called shadow factor). 

\begin{figure}
\centering
\includegraphics[width=3.5in]{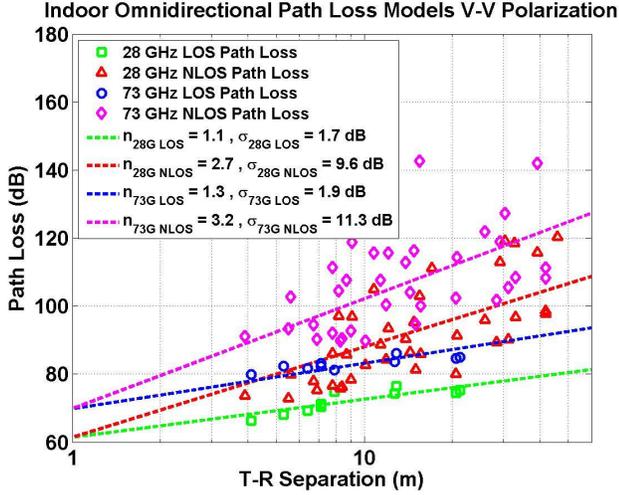}
\caption{28 GHz and 73 GHz omnidirectional close-in reference distance path loss models with respect to a 1 m free space reference distance for co-polarized antennas using 15 dBi and 20 dBi (30$^{\circ}$ and 15$^{\circ}$ HPBW) TX and RX antenna pairs, respectively, from data measured in a typical office indoor environment.}
\label{fig:OmniVV}
\end{figure}

\begin{figure}
\centering
\includegraphics[width=3.5in]{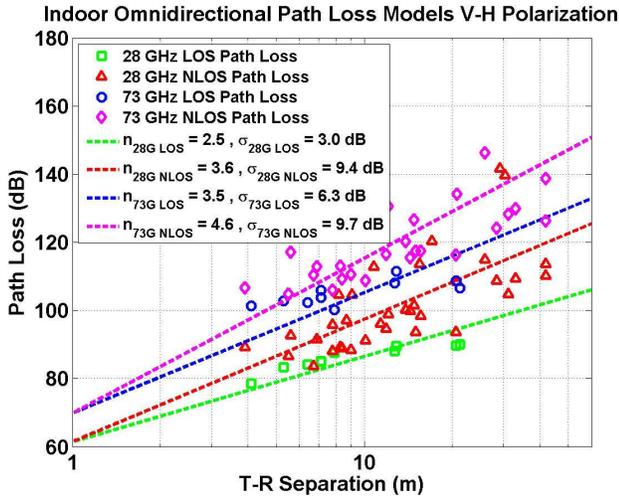}
\caption{28 GHz and 73 GHz omnidirectional close-in reference distance path loss models with respect to a 1 m free space reference distance for cross-polarized antennas using 15 dBi and 20 dBi (30$^{\circ}$ and 15$^{\circ}$ HPBW) TX and RX antenna pairs, respectively, from data measured in a typical office indoor environment.}
\label{fig:OmniVH}
\end{figure}

Fig.~\ref{fig:OmniVV} and Fig.~\ref{fig:OmniVH} show the 28 GHz and 73 GHz LOS and NLOS omnidirectional path loss data sets for V-V and V-H polarization scenarios, respectively, and corresponding 1 m close-in free space reference distance mean path loss equation lines. The LOS PLEs and shadowing factors for the V-V scenario were measured to be 1.1 and 1.7 dB, and 1.3 and 1.9 dB, at 28 GHz 73 GHz, respectively, which is slightly better than free space propagation ($n = 2$), occuring from the waveguide effect, where propagation multipath components are guided along hallways and constructively interfere at the RX antenna. The NLOS path loss exponents were measured to be 2.7 and 3.2 at 28 GHz and 73 GHz in the co-polarization scenario, respectively. In the cross-polarization scenario, the LOS path loss exponents were measured to be 2.5 and 3.5 at 28 GHz and 73 GHz, respectively, providing a cross-polarization discrimination (XPD) ratio of 14 dB and 23 dB per decade over the co-polarization scenario. Co-polarized antennas must be employed in a LOS environments to avoid signal degradation due to polarization mismatch~\cite{Maltsev60G}.

 \section{Millimeter-Wave Channel Time Dispersion Properties}

The time dispersion properties of wideband channels are generally characterized using the RMS delay spread, which describes the multipath time dispersion and coherence bandwidth nature of the channel that can be used to estimate data rates and bandwidth limitations for multipath channels~\cite{TedBook}. The RMS delay spread is the square root of the second central moment of the power delay profile, defined as:

\begin{equation} \label {eq:RMS}
\sigma_{\tau} = \sqrt{\overline{{\tau}^2}-(\bar\tau)^2 }
\end{equation}

where, 

\begin{equation} \label {eq:MeanED}
\bar\tau = \frac{\sum_{k}P(\tau_{k}) \tau_{k}} {\sum_{k}P(\tau_{k})}
\end{equation}

\begin{equation} \label {eq:SCM}
\overline{\tau^2} = \frac{\sum_{k}P(\tau_{k}) {\tau_{k}}^2} {\sum_{k}P(\tau_{k})}
\end{equation}

\noindent where $P(\tau_k)$ is the measured power in mW in time delay bin $\tau_k$.

\begin{figure}
\centering
\includegraphics[width=3.5in]{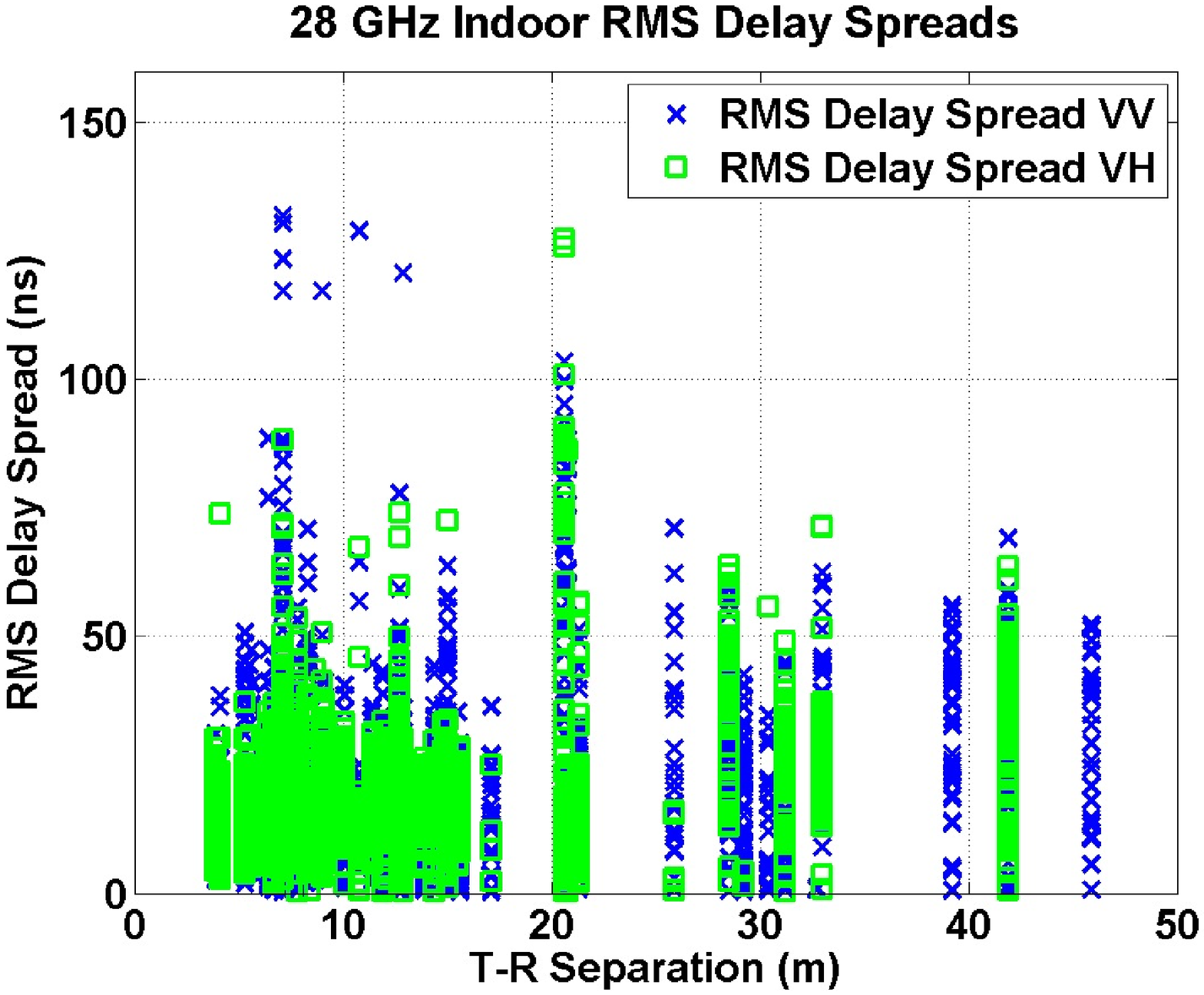}
\caption{28 GHz indoor RMS delay spread as a function of T-R separation distance for V-V and V-H polarization using a pair of 15 dBi gain (30$^{\circ}$ HPBW) antennas. V-V means vertically-polarized antenna at both the TX and RX, and V-H means vertically-polarized antenna at the TX and horizontally-polarized antenna at the RX.}
\label{fig:RMS_28}
\end{figure}

\begin{figure}
\centering
\includegraphics[width=3.5in]{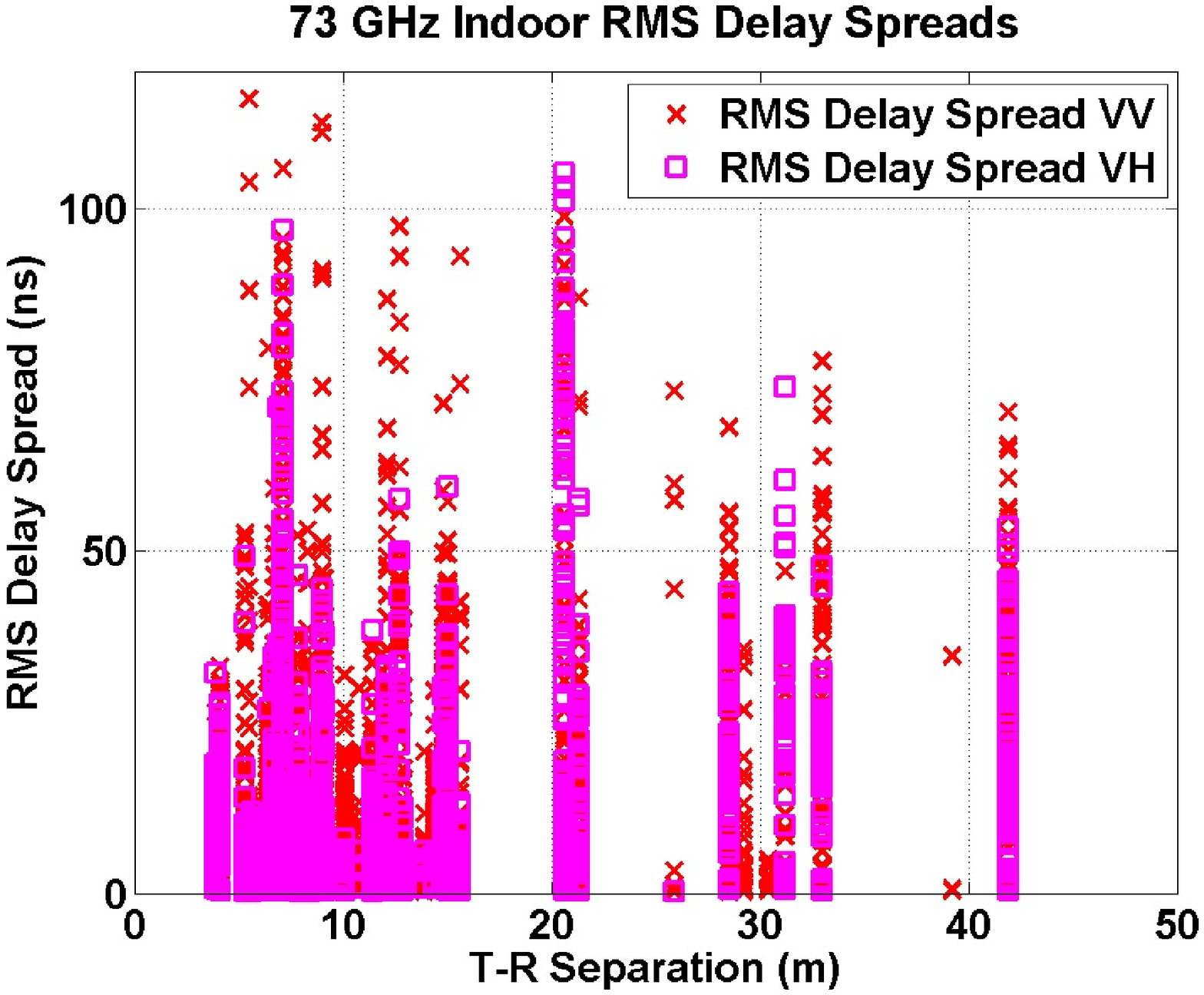}
\caption{73 GHz indoor RMS delay spread as a function of T-R separation distance for V-V and V-H polarization using a pair of 20 dBi gain (15$^{\circ}$ HPBW) antennas. V-V means vertically-polarized antenna at both the TX and RX, and V-H means vertically-polarized antenna at the TX and horizontally-polarized antenna at the RX.}
\label{fig:RMS_73}
\end{figure}

\begin{figure}
\centering
\includegraphics[width=3.5in]{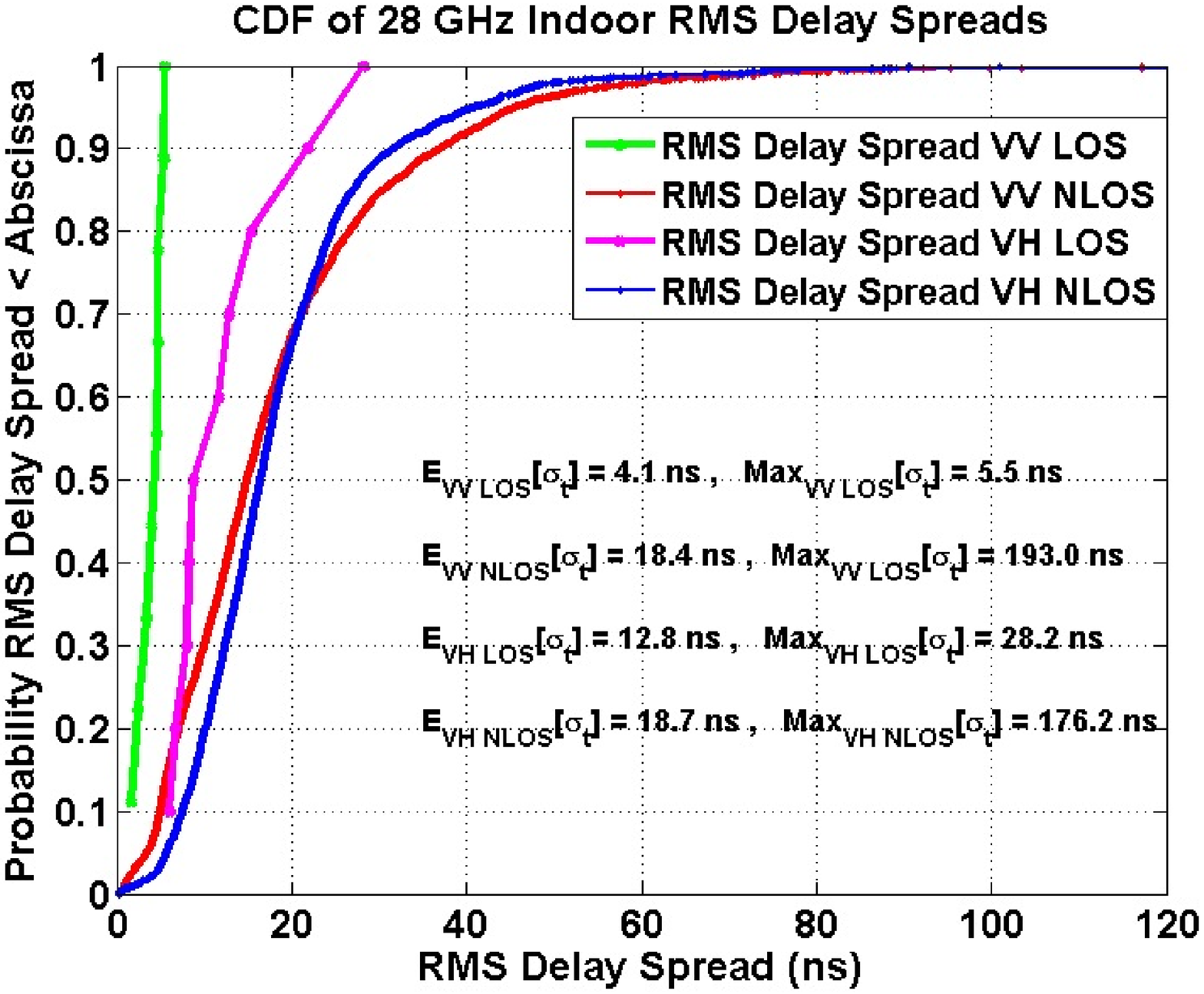}
\caption{28 GHz indoor RMS delay spread CDF with T-R separation distances ranging from 3.9 m to 45.9 m in a typical office environment (including LOS and NLOS environments) for V-V and V-H polarization scenarios using a pair of 15 dBi gain (30$^{\circ}$ HPBW) antennas.}
\label{fig:CDF_28}
\end{figure}

\begin{figure}
\centering
\includegraphics[width=3.5in]{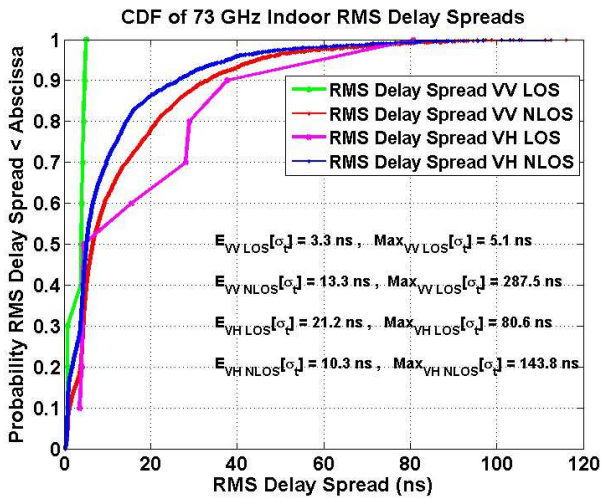}
\caption{28 GHz indoor RMS delay spread CDF with T-R separation distances ranging from 3.9 m to 45.9 m in a typical office environment (including LOS and NLOS environments) for V-V and V-H polarization scenarios using a pair of 20 dBi gain (15$^{\circ}$ HPBW) antennas.}
\label{fig:CDF_73}
\end{figure}

 Fig.~\ref{fig:RMS_28} and Fig.~\ref{fig:RMS_73} show the 28 GHz and 73 GHz directional RMS delay spreads as a function of T-R separation distance, respectively, for both co- and cross-polarized scenarios, indicating little correlation over distance. Fig.~\ref{fig:CDF_28} and Fig.~\ref{fig:CDF_73} show the cumulative distribution function (CDF) of directional RMS delay spreads at 28 GHz and 73 GHz for different antenna polarization combinations and environment scenarios. In LOS boresight scenario, the mean RMS delay spreads were 4.1 ns and 3.3 ns at 28 GHz and 73 GHz, respectively, indicating that most of received power is contained in the first arriving LOS multipath component. In NLOS, the mean RMS delay spreads were measured to be 18.4 ns and 13.3 ns when considering all unique pointing angles that were not aligned on boresight in LOS and NLOS environments. Table~\ref{tbl:RMS_CDF} provides the mean, standard deviation, and maximum values of directional RMS delay spread statistics at 28 GHz and 73 GHz, in both co- and cross-polarized scenarios.
 
 
RMS delay spreads CDF at 28 GHz are shown in Fig.~\ref{fig:CDF_28}, 90\% of the arriving angles are within 5.5 ns and 21.8 ns for co- and cross-polarized scenarios in LOS environments, respectively, whereas in NLOS environments, 90\% of the arriving angles are within 36.4 ns and 31.4 ns for co- and cross-polarized scenarios, respectively. RMS delay spreads CDF at 73 GHz are shown in Fig.~\ref{fig:CDF_73}, 90\% of the arriving angles are within about 5.1 ns and 37.8 ns for co- and cross-polarized scenarios in LOS environments, respectively, whereas in NLOS environments, 90\% of the arriving angles are within 33.2 ns and 26.0 ns for co- and cross-polarized scenarios, respectively. For the V-V scenario, the mean RMS delay spreads are 4.1 ns and 3.3 ns in the LOS environment, at 28 GHz and 73 GHz, respectively, which are less than 18.4 ns and 13.3 ns in the NLOS environment. The LOS components with short delays lead to reduced RMS delay spread in LOS environments, while the strong reflected components (relative to the LOS path) with long delays contribute significantly to large RMS delay spread in the NLOS environments \cite{Geng60G}. For the V-H scenario, the mean RMS delay spreads are 21.2 ns at 73 GHz in the LOS environment, which are larger than 10.3 ns in the NLOS environment, since polarization mismatch weaken the LOS component, while reflections (which may change the polarization of the signal) enhance the reflected components, resulting in increased RMS delay spread.

 The work in \cite{ShuMIMOMag} suggests a simple algorithm to find the best beam directions that can simultaneously minimize both RMS delay spread and path loss (finding the best paths for both maximum SNR and very simple equalization). By selecting a beam with both low RMS delay spread and path loss, relatively high power can be received using directional antennas without complicated equalization, meaning that low latency single carrier (wideband) modulations may be a viable candidate for future mmWave wireless communications \cite{Ghosh2014JASC}. The measured values presented in this paper can help in implementing beamforming or beam finding algorithms to systemically search for the strongest TX and RX pointing angles that result in the lowest path loss or link attenuation.


\begin{table}
\caption{Comparison of mean RMS delay spread, standard deviation and maximum RMS delay spread at 28 GHz and 73 GHz for co- and cross-polarization combinations and in LOS and NLOS scenarios using arbitrary unique point angle environment with high directional horn antennas.}
  \label{tbl:RMS_CDF}
 \centering
 \begin{center}
  \begin{tabular}{|c|c|c|c|c|c|c|}
	
	 \hline
	\multirow{2}{*}{ \textbf{Scenario} }&\multicolumn{3}{c|}{\textbf{28 GHz}}  & \multicolumn{3}{c|}{\textbf{73 GHz}}   \\ \cline{2-7}
	                                                     &\textbf{Mean} & \textbf{Std.} & \textbf{Max.} & \textbf{Mean} & \textbf{Std.}& \textbf{Max.} \\ \hline                                                                                
	  \textbf{LOS V-V}     & 4.1  & 1.3   & 5.5  &  3.3  & 1.8   & 5.1  \\  \hline
	  \textbf{NLOS V-V}    & 18.4  & 14.9  & 193.0  &  13.3  & 16.2   & 287.5  \\  \hline
	  \textbf{LOS V-H}     & 12.8  & 7.2  & 125.9  &  21.2  & 13.9   & 80.6  \\  \hline
	  \textbf{NLOS V-H}    & 18.7  & 12.4  & 176.2  &  10.3   & 10.3   & 143.8  \\  \hline                                                                           
	                                    
 \end{tabular}
\end{center}
\end{table}

 \section{Conclusion }\label{sec:conclustion}
 
 This paper presented 28 GHz and 73 GHz indoor propagation directional and omnidirectional path loss models obtained from two extensive ultrawideband propagation measurement campaigns using a sliding correlator channel sounder and high-gain directional horn antennas. In LOS, the omnidirectional path loss exponents were measured to be 1.1 and 1.3 with respect to a 1 m close-in free space reference distance, indicating significant improvements over free space propagation as a result of the constructive intereference of multipath signals. In NLOS, we measured omnidirectional path loss exponents of 2.7 and 3.2 at 28 GHz and 73 GHz, respectively, showing increased signal attenuation over distance resulting from obstructions between the TX and RX. The directional path loss exponents were measured to be 4.5 and 5.1 at 28 GHz and 73 GHz, respectively, when considering arbitrary unique pointing angles, but were decreased to 3.0 and 4.3 when searching for the strongest TX-RX angle pointing link at each RX location, showing great value in beamforming at the base station and mobile handset for SNR enhancement and increase coverage. The mean and maximum RMS delay spread values were found to be 18.4 ns and 193.0, and13.3 ns and 287.5 ns at 28 GHz and 73 GHz in LOS and NLOS environments, indicating that strong multipath components can arrive at large time delays. The channel models presented here can be used for mmWave system-wide simulations and radio-system design in indoor environments for next generation 5G communication systems.  
 \\
 \\

\bibliographystyle{IEEEtran}
\bibliography{ICC2015}
\end{document}